\documentclass[aps,prl,reprint,groupedaddress]{revtex4-2}
\usepackage{color}
\usepackage{multirow}
\usepackage{graphicx}
\newcommand{\BB}{\begin{equation}}
\newcommand{\BE}{\end{equation}}
\begin{document}
%\begin{CJK*}{GB}{} % 
% Use the \preprint command to place your local institutional report
% number in the upper righthand corner of the title page in preprint mode.
% Multiple \preprint commands are allowed.
% Use the 'preprintnumbers' class option to override journal defaults
% to display numbers if necessary
%\preprint{}

%Title of paper
\title{
Representation of degree correlation using eigenvalue decomposition and its application to epidemic models
}

% repeat the \author . \affiliation  etc. as needed
% \email, \thanks, \homepage, \altaffiliation all apply to the current
% author. Explanatory text should go in the []'s, actual e-mail
% address or url should go in the {}'s for \email and \homepage.
% Please use the appropriate macro foreach each type of information

% \affiliation command applies to all authors since the last
% \affiliation command. The \affiliation command should follow the
% other information
% \affiliation can be followed by \email, \homepage, \thanks as well.
\author{Satoru Morita}
\email[]{morita.satoru@shizuoka.ac.jp}
%\homepage[]{Your web page}
%\thanks{}
%\altaffiliation{}
\affiliation{Department of Mathematical and Systems Engineering, Shizuoka University, Hamamatsu 432-8561, Japan}

%Collaboration name if desired (requires use of superscriptaddress
%option in \documentclass). \noaffiliation is required (may also be
%used with the \author command).
%\collaboration can be followed by \email, \homepage, \thanks as well.
%\collaboration{}
%\noaffiliation

\date{\today}

\begin{abstract}
Degree correlation plays a crucial role in studying network structures; however, its varied forms pose challenges to understanding its impact on network dynamics. This study devised a method that uses eigenvalue decomposition to characterize degree correlations. Additionally, the applicability of this method was demonstrated by approximating the basic and type reproduction numbers in an epidemic network model. The findings elucidate the interplay between degree correlations and epidemic behavior, thus contributing to a deeper understanding of complex networks and their dynamics.
\end{abstract}

% insert suggested keywords - APS authors don't need to do this
%\keywords{}

%\maketitle must follow title, authors, abstract, and keywords
\maketitle
%\end{CJK*}

Network science provides a powerful tool for understanding complex systems, in which entities are represented as interconnected nodes and their relationships as links \cite{albert,newman2006,dorogovtsev,barabasi2016,newman2018}.
This tool has aided the elucidation of the structure and dynamics of various systems, such as social networks, biological interactions, and the Internet. 
The study of network structures, such as degree distributions, degree correlations, and clustering, has revealed universal principles having interdisciplinary applicability.
By leveraging insights from network science, real-world systems can be better understood and effective strategies and interventions can be developed.

This study focused on degree correlation, which refers to the relationship between the degrees of connected nodes in a network.
The degree of a node represents the number of connections it contains, and the degree correlation investigates whether nodes with similar degrees tend to be connected (positive correlation or assortativity) or nodes with different degrees tend to be connected (negative correlation or disassortativity).
However, degree correlations are complex and cannot be divided based on whether they are positive or negative.
In particular, when the degree distribution has a fat tail, restrictions on the links between hubs cause structural disassortativity \cite{barabasi2016,park2003}.
Therefore, the effect of the degree of correlation on the spread of infectious diseases is not necessarily simple and remains unclear.

The characteristics of degree correlation are herein expressed using eigenvalue decomposition. 
Some studies have performed eigenvalue decomposition of adjacency matrices \cite{mieghem2011,ferraz2018}. 
However, to the author's best knowledge, no prior study has focused on eigenvalue decomposition for matrices that represent degree correlation, as was done in this study.
This method was applied to an epidemic network model to approximate the basic and type reproduction numbers. 
%We also discuss the scope of application of this approximation.

The degree distribution $p_k$ is defined as the probability that a randomly selected node has degree $k$.
The probability distribution of the degree at one end of a randomly selected link is called the excess degree and is given by
\begin{equation}
q_k =k p_k/{\langle k \rangle},
\end{equation}
where $\langle k \rangle=\sum_k k p_k$.
It must be noted that $p_k$ and $q_k$ constitute $k_{max}$-dimensional vectors if the maximum degree is $k_{max}$ and isolated nodes (with degree zero) are excluded.
Let $e_{hk}$ be the probability that the two ends of a randomly selected link are nodes of degrees $h$ and $k$ \cite{barabasi2016,newman2018}.
If each degree is independent, 
\begin{equation}
e_{hk}=q_h q_k.
\end{equation}
The eigenvalue decomposition of matrix $e_{hk}-q_h q_k$ yields the following equation:
\begin{equation}
e_{hk}=q_h q_k+\sum_{i=1}\lambda_i f_h^{(i)} f_k^{(i)},
\label{eq_ehk}
\end{equation}
where $\lambda_i$ is the eigenvalue of the symmetric matrix $e_{hk}-q_h q_k$ on the order $|\lambda_1|\geq|\lambda_2|\geq|\lambda_3|\cdots$ and $f^{(i)}$ are the corresponding eigenvectors.
The summation in Eq.~(\ref{eq_ehk}) over the rank of matrix $e_{hk}-q_h q_k$ yields the exact expression, whereas stopping the summation at the appropriate point yields an approximate formula.

The degree correlation was described using the Pearson correlation coefficient between the degrees at the two ends of the same link:
\begin{equation}
r=\frac{\sum_{h}\sum_{k}hk(e_{hk}-q_h q_{k})}
{\sum_{k}k^2 q_k-(\sum_{k}k q_k)^2}.
\end{equation}
This coefficient is also referred to as the assortativity coefficient \cite{newman2002}.
By substituting Eq.~(\ref{eq_ehk}), the following results:
\begin{equation}
r=\displaystyle \frac{\langle k \rangle^2}
{\langle k^3 \rangle\langle k \rangle-\langle k^2 \rangle^2}\sum_{i=1} \lambda_i \langle k f_k^{(i)} \rangle^2,
\label{eq_r}
\end{equation}
where the following notation is used
\begin{equation}
\langle k f_k^{(i)} \rangle=\sum_{k=1}^{k_{max}}  k f_k^{(i)}.
\end{equation}
Thus, if $|\lambda_1|\gg|\lambda_2|$, the positivity or negativity of $\lambda_1$ determines the positivity or negativity of $r$.
To characterize the degree correlation in more detail,
the average degree of the nearest neighbors $k_{\mbox{\tiny nn}}(k)$ is used \cite{pastor2001x}. 
Using Eq.~(\ref{eq_ehk}), it can be calculated as
\begin{equation}
k_{\mbox{\tiny nn}}(k)=\sum_{h=1}^{k_{max}} h e_{hk}/q_k =
\frac{\langle k^2\rangle}{\langle k\rangle}
+\sum_{i=1} \lambda_i  \langle k f_k^{(i)} \rangle \frac{f_k^{(i)}}{q_k}.
\label{eq_knnk}
\end{equation}

\begin{figure}[t]
\begin{center}
\includegraphics[width=\linewidth]{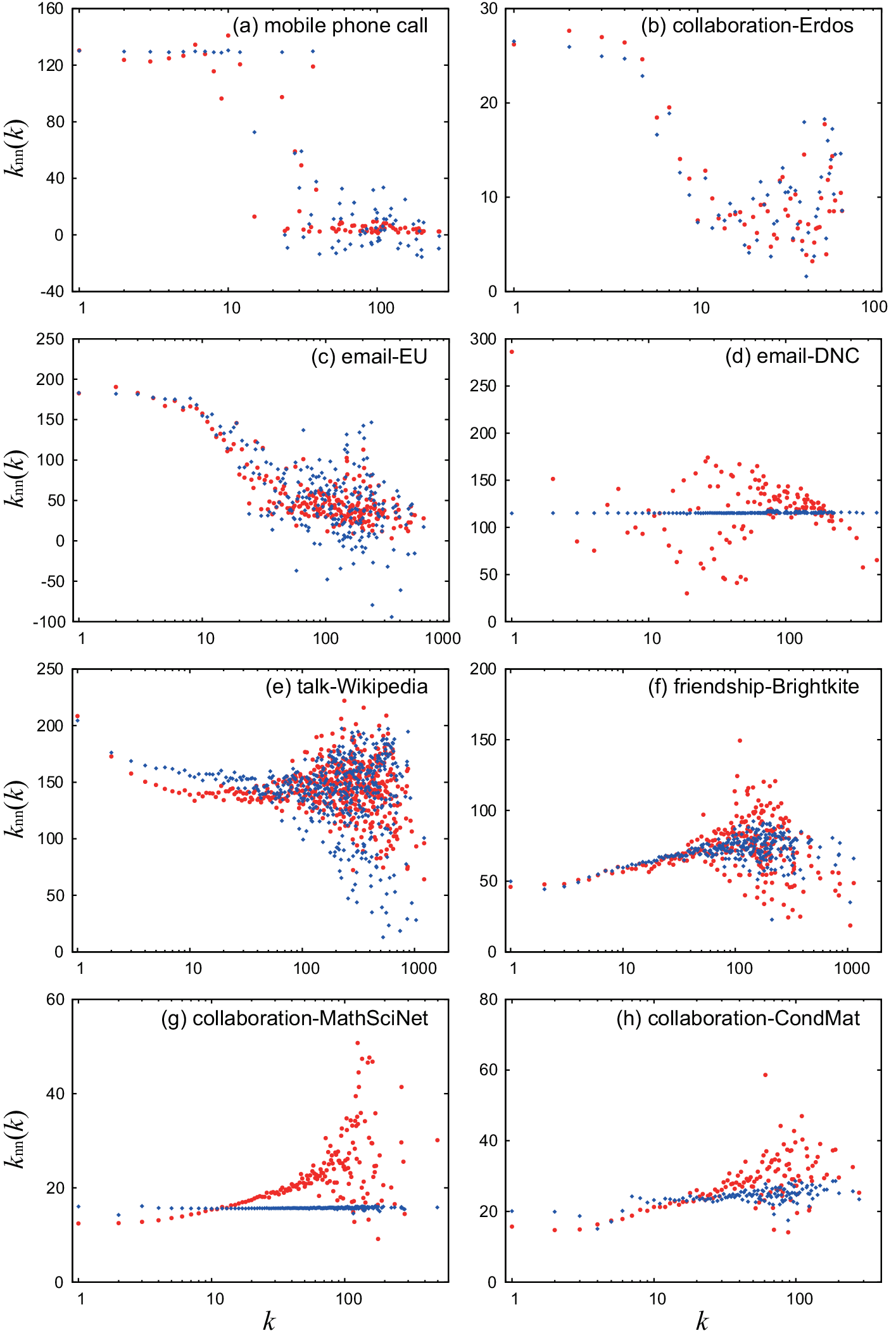}\\%
\caption{
Average degrees of the nearest neighbors $k_{\mbox{\tiny nn}}(k)$ (red) for the eight actual networks and the first-order approximation (blue) based on eigenvalue decomposition, as shown in Eq.~(\ref{eq_knnk}). \label{f2}}
\end{center}
\end{figure}

\begin{table*}[hbtp]
\caption{Examples of several real networks sorted in ascending order with respect to the degree correlation coefficient $r$: (a) personal mobile phone call network among a small set of core users at the Massachusetts Institute of Technology \cite{nr,eagle2006reality},
(b) coauthorship network among mathematicians who coauthored papers with Erd\"{o}s (Erd\"{o}s number is one) or with those mathematicians (Erd\"{o}s number is two) \cite{nr,batagelj2000some},  
(c) network of email communications at a large European research institute \cite{Leskovec},
(d) network of email communications at the Democratic National Committee(DNC) \cite{nr},
(e) communication networks on Talk Page in Wikipedia \cite{nr,leskovec2010predicting},
(f) friendship networks in Brightkite, which was once a location-based social networking service \cite{friendship},
(g) coauthorship network among mathematicians on mathematical review collection of the American Mathematical Society (MathSciNet) \cite{nr,palla2008fundamental}, and
(h) coauthorship network among physicists on Arxiv cond-mat (condensed matter physics) \cite{nr,condmat}.
Each column shows the number of nodes, number of links, average degree, maximum degree, matrix rank
 $e_{hk}-q_h q_k$, degree correlation coefficient $r$, average clustering coefficient $C$, and the first and second
 eigenvalues of the matrix $e_{hk}-q_h q_k$.
\label{networks}}
\centering
\begin{tabular}{lrrrrrrcrr}
\hline
Network & Nodes & Edges & $\langle k \rangle$ & $k_{max}$ &Rank&  $r$ \ \ & $C$ & $\lambda_1$ \hspace{2mm} &$\lambda_2$ \hspace{2mm} \\
\hline
(a) Mobile phone call \cite{nr,eagle2006reality} &
6809 & 7680 & 2.3 & 261 & 74 & -0.68 & 0.02 &  -0.17525 &   0.00357 \\
(b) Collaboration---Erd\"{o}s \cite{nr,batagelj2000some}&
5094 & 7515 & 3.0 & 61 & 56 & -0.44 & 0.07 &     -0.07094 & -0.00437\\
(c) Email---EU \cite{Leskovec} &
32430 & 54397 & 3.4 & 623 & 232& -0.38 & 0.11& -0.06209 &   0.00230\\
%(d) email-Enron \cite{enron,enron2}&
%36692 & 367662&20.0 & 2766 & 333 &-0.11 & 0.50& 0.01094 &   0.00724\\
(d) Email---DNC \cite{nr}&
906 & 12085 & 26.6 & 462 & 134 & -0.09 & 0.61 &   0.02199 &   0.01391\\
(e) Talk---Wikipedia \cite{nr,leskovec2010predicting} &
92117 & 360767 & 7.8 & 1220 & 485 & -0.03&0.06&-0.00603 &   0.00062\\
(f) Friendship---Brightkite \cite{friendship} &
58228 & 428156 & 7.4 & 1134 & 261 & 0.01 & 0.17& 0.00555 & -0.00248\\
(g) Collaboration---MathSciNet \cite{nr,palla2008fundamental} &
391529 & 873775 & 4.5 & 496 &162 & 0.12	&0.40& 0.01581 & 0.01291\\
(h) Collaboration---CondMat \cite{nr,condmat} &
21363 & 91286 & 8.5 & 279 &122 &  0.13 & 0.63& 0.00775 &  0.00761\\
\hline
\end{tabular}
\end{table*}
\begin{figure}[tb]
\begin{center}
\includegraphics[width=\linewidth]{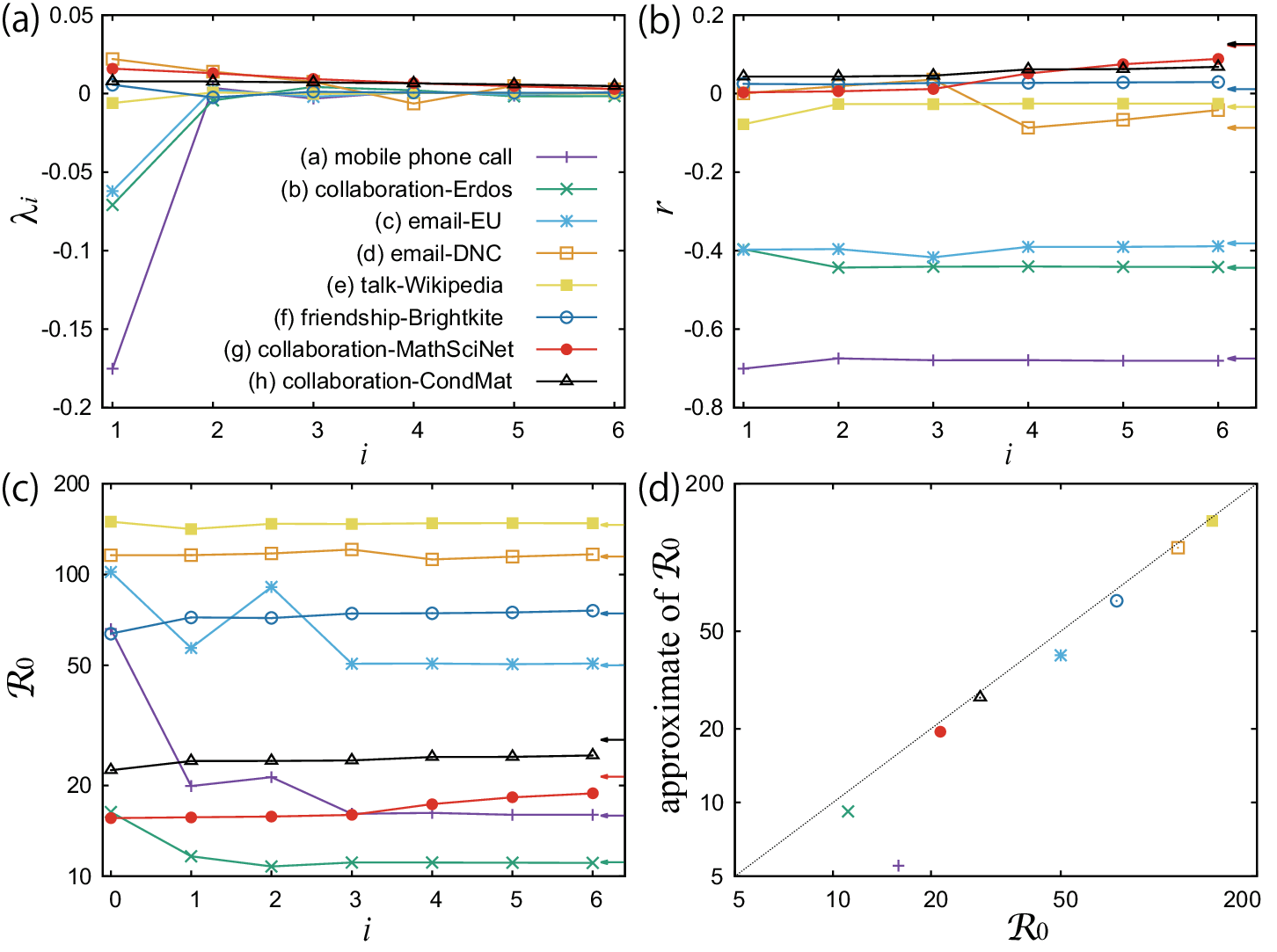}\\%
\caption{
Eigenvalue decomposition results for the eight real networks shown in Table ~\ref{networks}. 
(a) Eigenvalues of the matrix $e_{hk}-q_h q_kz$ with the six largest absolute values.
(b) Degree correlation coefficient $r$ considered up to the $i$th order in the eigenvalue decomposition given by Eq.~(\ref{eq_r}).
The arrows on the right edge of the panel indicate the actual value of $r$.
(c) The value of the basic reproduction number $\mathcal{R}_{0}$ is calculated based on Eq.~(\ref{r0a}) when $\beta =1$.
The arrows on the right edge indicate the value of $\mathcal{R}_{0}$ given by the spectral radius of Eq.~(\ref{eq_gi}).
For $i=0$, there is no degree correlation and $\mathcal{R}_{0}=\langle k^2\rangle/\langle k\rangle$.
(d) Relationship between the actual $\mathcal{R}_{0}$ and the $\mathcal{R}_{0}$ given by the approximation of Eq.~(\ref{r0b}) when $\beta = 1$. \label{f1}}
\end{center}
\end{figure}
The eigenvalue decomposition (as aforementioned) was adopted in the eight empirical social networks shown in Table ~\ref{networks}.
Fig.~\ref{f1}(a) shows the eigenvalues up to the sixth order of the absolute value (the first and second eigenvalues are shown in Table ~\ref{networks}).
The first eigenvalue is prominent for networks (a), (b), and (c), which have strong, negative-degree correlations and small clustering coefficients. 
However, in the other four cases, except for (f), the first and second eigenvalues have almost absolute values, thus suggesting that the first eigencomponent alone cannot describe the degree correlation.
Fig.~\ref{f1}(b) shows the behavior of the degree correlation coefficient $r$ when the eigenvalue expansion is approximated to the $i$th order.
Again, in networks (a), (b), and (c), the negative correlation was well reproduced with a few components, whereas in the other networks, it was not as good.
Fig.~\ref{f2} presents a comparison of $k_{\mbox{\tiny nn}}(k)$ when only the first eigencomponent is considered with the actual $k_{\mbox{\tiny nn}}(k)$ (the former is shown in blue, and the latter in red).
In cases (a), (b), (c), and (f), the approximation reproduced the tendency of the degree correlation well, whereas in cases (d) and (g), it did not.
%\begin{equation}
%\langle f \rangle := \sum_{i=1}^{k_{max}}f_k^{(i)}p_k=0
%\end{equation}
%to ensure that the marginal distributions $\sum_h e_{hk}$ and $\sum_h e_{kh}$ coincide with $q_k$. 
%\begin{equation}
%p_{k|h}=e_{kh}/q_h
%\end{equation}
%\begin{equation}
%k_{nn}(k)= \frac{\langle k^2\rangle}{\langle k\rangle}+\varepsilon \frac{\langle k f(k)\rangle}{\langle k\rangle}
%\frac{f(k)}{k}
%\end{equation}

Let us now examine the effect of degree correlations on diffusion phenomena in the network.
The following epidemic model was considered: 
\begin{equation}
 \frac{d\rho_k(t)}{dt}=-\rho_k(t)+\beta k
[1-\rho_k(t)]\sum_{h=1}^{k_{max}}{k}p_{h|k}\rho_{h}(t),
\label{eq2}
\end{equation}
where $\rho_k(t)$ represents the density of infected
nodes within degree class $k$, and
$p_{h|k}$ is the conditional probability that a node of degree $k$ is connected to a node of degree $h$ and is given by $p_{h|k}=e_{hk}/q_k$.
Herein, $\beta$ is the transmission rate, and the time scale is set such that the recovery rate is unity.
Eq.~(\ref{eq2}) was proposed as a degree-based mean-field approximation of epidemic models on fat-tailed networks with a few loops \cite{boguna2002,boguna2003,moreno,morita2016,morita2022,RMP2015,kiss2017}.

To measure the transmission potential, the basic reproduction number $\mathcal{R}_{0}$ has been used; this is the average number of secondary infections caused by a typical infection in a
completely susceptible population \cite{anderson,diekmann}. 
When $\mathcal{R}_{0}>1$, the infection can spread throughout the host population, whereas when $\mathcal{R}_{0}<1$, the infection
cannot spread. 
The value of $\mathcal{R}_{0}$ is given by the spectral radius of the next-generation matrix $G$.
For Eq.~(\ref{eq2}), the component $G_{kh}$ of the next-generation matrix is expressed as
\begin{eqnarray}
G_{kh}&=&\beta k p_{h|k}\nonumber\\
&=&\beta \left(k q_h+\sum_{i=1} \lambda_i k f_h^{(i)}f_k^{(i)}/q_k \right),
\end{eqnarray}
which is the average number of secondary cases arising from a node of degree $h$ to nodes of degree $k$ \cite{diekmann2010,morita2021}.
If the following vertical and horizontal vectors are represented using blas and kets
 \begin{eqnarray}
|k\rangle &=&(1,2,\dots,k_{max})^T,\\
\langle k| &=&(q_1,q_2,\dots,q_{k_{max}}),\\
|f^{(i)}\rangle &=&\left(\frac{f_1^{(i)}}{q_1},\frac{2 f_2^{(i)}}{q_2},\dots,\frac{k_{max}f_{k_{max}}^{(i)}}{q_{k_{max}}}\right)^T,\\
\langle f^{(i)} |&=&(f_1^{(i)},f_2^{(i)},\dots,f_{k_{max}}^{(i)}),
\end{eqnarray}
the next-generation matrix can be written as
\begin{equation}
G=\beta\left(|k\rangle \langle k|+\sum_{i=1} \lambda_i |f^{(i)}\rangle \langle f^{(i)}|\right).
\label{mainresult}
\end{equation}
Because the eigenvectors $G$ are linearly coupled and consist of $|k\rangle$ and $|f^{(i)}\rangle$,
the eigenvalue problem for $G$ is replaced by the eigenvalue problem for the ($i+1$)-dimensional matrix as follows:
\begin{equation}
G^{(i)}=\beta\left(
\begin{array}{cccc}
\langle k|k\rangle &\langle k|f^{(1)}\rangle & \cdots &\langle k|f^{(i)}\rangle\\
\lambda_1 \langle f^{(1)}|k\rangle &\lambda_1\langle f^{(1)}|f^{(1)}\rangle & \cdots &\lambda_1\langle f^{(1)}|f^{(i)}\rangle\\
\vdots & \vdots & \ddots &\vdots \\
\lambda_i \langle f^{(i)}|k\rangle &\lambda_i \langle f^{(i)}|f^{(1)}\rangle & \cdots &\lambda_i \langle f^{(i)}|f^{(i)}\rangle
\end{array}\right).
\label{eq_gi}
\end{equation}
When $i$ is the rank of $e_{hk}$, Eq.~(\ref{eq_gi}) is the exact expression, whereas for small $i$, this is an approximate expression.

Alternatively, if $|\lambda_i|\ll 1$ for $i=1,2,\dots$, then another approximation is possible.
Neglecting the terms above the second order of $\lambda_i$, an approximation of the basic reproduction number is obtained as
\begin{equation}
\mathcal{R}_{0}\simeq \beta\left(\langle k|k\rangle
+\sum_{i=1} \lambda_i\frac{\langle k|f^{(i)}\rangle \langle f^{(i)}|k\rangle}{\langle k|k\rangle}\right).
%\mathcal{R}_{0}\simeq \beta\left(\frac{\langle k^2\rangle}{\langle k\rangle}
%+\sum_{i=1} \lambda_{i=1}\frac{\langle k\rangle\langle kf_k^{(i)}\rangle^2}{\langle k^2\rangle}\right).
\label{r0a}
\end{equation}
%If $1\gg|\lambda_1|\gg|\lambda_2|$, 
%then ignoring terms after $\lambda_2$ and computing the dominant eigenvalue of $G$ in an approximation to the first order of 
%$\lambda_1$ yields the approximation for the basic reproduction number:
%\begin{equation}
%\mathcal{R}_{0}\simeq \beta\left(\frac{\langle k^2\rangle}{\langle k\rangle}
%+\lambda_1\frac{\langle k\rangle\langle kf_k^{(1)}\rangle^2}{\langle k^2\rangle}\right).
%\end{equation}
Considering 
$\langle k|k\rangle =\langle k^2\rangle/\langle k\rangle$  
$\langle k|f^{(i)}\rangle = \langle f^{(i)}|k\rangle = \langle k f_k^{(i)} \rangle$, and Eq.~(\ref{eq_r}),
%\langle f^{(i)}|f^{(j)}\rangle &=& \sum_{k=1}^{k_{max}}k f_k^{(i)}f_k^{(j)}/q_k.
Eq.~(\ref{r0a}) can be rewritten as follows:
\begin{equation}
\mathcal{R}_{0}\simeq \beta\frac{\langle k^2\rangle}{\langle k\rangle}\left[1
+r\left(\frac{\langle k^3\rangle \langle k\rangle}{\langle k^2\rangle^2}-1\right)\right].
\label{r0b}
\end{equation}
Fig.~\ref{f1}(d) presents a comparison of Eq.~(\ref{r0b}) and the exact expression using the spectral radius of Eq.~(\ref{eq_gi}),
and it shows that Eq.~(\ref{r0b}) is a good approximation except for networks (a), (b), and (c), which yield strong, negative-degree correlations.
%Eq.~(\ref{r0b}) is attractive in that it is given as a function of the degree correlation coefficient but may not have a good agreement for the real networks treated here (especially for (a) and (f)).

\begin{figure}[tb] 
\begin{center}
\includegraphics[width=\linewidth]{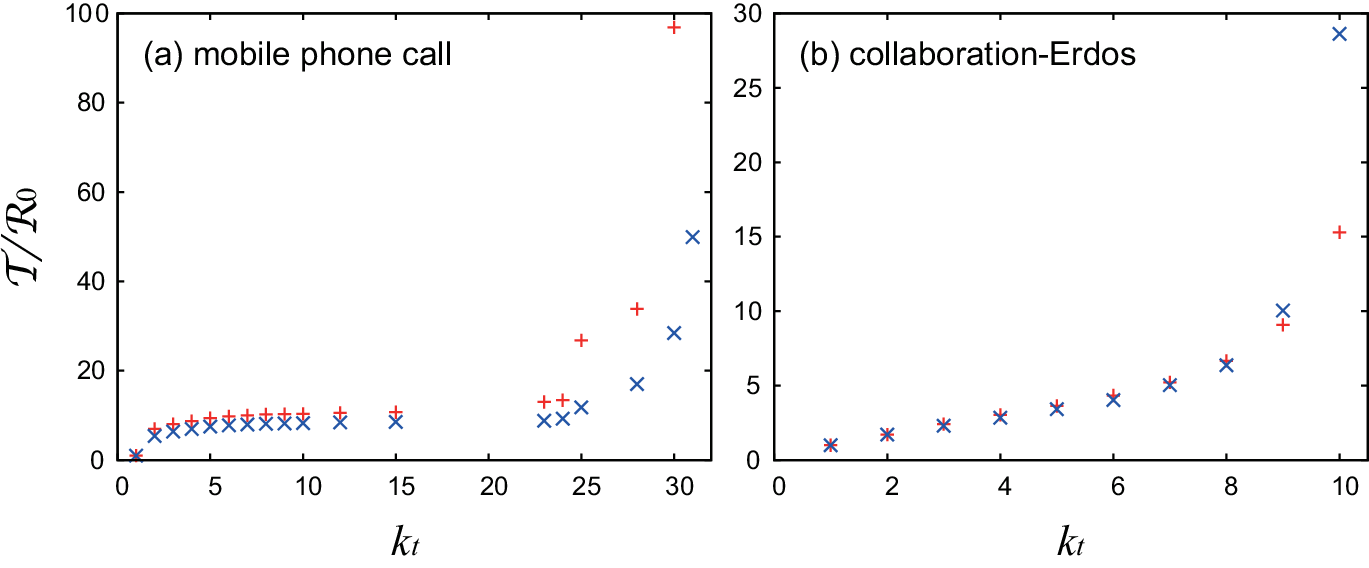}%
\caption{
(Color online)
Type reproduction number $\mathcal{T}$ (for two real networks (a) and (b)) when targeting nodes with degrees greater than or equal to $k_t$.
Red plus ``$+$'' represents the result obtained from Eq.~(\ref{eq_t1}) and blue cross ``$\times$'' represents the approximation from Eq.~(\ref{eq_t2}) with $i=1$. 
When $\mathcal{T}$ is finitely defined, immunizing only the target can suppress the spread of infection.
This limit is independent of $\beta$, with $k_t\leq31$ for (a) ($k_t\leq30$ in the approximate calculation) and $k_t\leq10$ for (b) (also in the approximate calculation).
\label{f3}}
\end{center}
\end{figure}

Finally, a method was devised to calculate the type reproduction number $\mathcal{T}$, which is used to 
to explore the effect of immunizing a target group and is defined as the average number of secondary infections within the target group \cite{robert2003,heesterbeek2007,morita2022}. 
If a proportion $1-1/\mathcal{T}$ of the target population is immunized, the spread of the infection can be suppressed.
Individuals with degree $k_t$ or greater are targeted.
The following matrix is introduced to calculate $\mathcal{T}$:
\begin{equation}
C=\beta\left(|k_t\rangle \langle k|+\sum_{i=1} \lambda_i |f_t^{(i)}\rangle \langle f^{(i)}|\right),
\end{equation}
where 
\begin{eqnarray}
%\langle k_t| &=&(0,\dots,0,q_{k_{t}},\dots,q_{k_{max}}),\\
%\langle k_<|       &=&(1,2,\dots,k_{t-1},0,\dots,0)\\
%\langle f_{t}^{(i)}| &=&(0,\dots,0,f_{k_t}^{(i)},\dots,f_{k_{max}}^{(i)}).
%\\ \langle f_<^{(i)}| &=&(f_1^{(i)},f_2^{(i)},\dots,f_{k_{t-1}}^{(i)},0,\dots,0)
|k_t\rangle &=&(0,\dots,0,k_t,\dots,k_{max})^T,\\
|f_t^{(i)}\rangle &=&\left(0,\dots,0,k_t\frac{f_{k_t}^{(i)}}{q_{k_t}},\dots,\frac{k_{max}f_{k_{max}}^{(i)}}{q_{k_{max}}}\right)^T.
\end{eqnarray}
Matrix $C$ consists of only those components of the next-generation matrix $G$ that infect the target population.
The type reproduction number is given by the spectral radius of the following matrix:
\begin{equation}
T=C(I-G+C)^{-1},
\label{eq_t1}
\end{equation}
where $I$ denotes the identity matrix \cite{robert2003,heesterbeek2007,morita2022}.
Because this eigenvalue problem can be simplified as before, the ($i+1$)-dimensional matrix is considered as follows:
\begin{equation}
C^{(i)}=\beta\left(
\begin{array}{cccc}
\langle k|k_t\rangle &\langle k|f_t^{(1)}\rangle & \cdots &\langle k|f_t^{(i)}\rangle\\
\lambda_1 \langle f^{(1)}|k_t\rangle &\lambda_1\langle f^{(1)}|f_t^{(1)}\rangle & \cdots &\lambda_1\langle f^{(1)}|f_t^{(i)}\rangle\\
\vdots & \vdots & \ddots &\vdots \\
\lambda_i \langle f^{(i)}|k_t\rangle &\lambda_i \langle f^{(i)}|f_t^{(1)}\rangle & \cdots &\lambda_i \langle f^{(i)}|f_t^{(i)}\rangle
\end{array}\right).
\label{eq_ci}
\end{equation}
The spectral radius of the following matrix was calculated:
\begin{equation}
T^{(i)}=C^{(i)}(I-G^{(i)}+C^{(i)})^{-1}.
\label{eq_t2}
\end{equation}
The type reproduction number $\mathcal{T}$ is defined only when the spectral radius of $G-C$ or $G^{(i)}-C^{(i)}$ is less than one:
Figure \ref{f3} shows (as an example) the effectiveness of this approximation method for two real networks.

In conclusion, this study demonstrated that degree correlation can be redescribed using eigenvalue decomposition and that the next-generation matrix can be represented as a matrix of a few orders.
This method can be used to obtain approximate values for basic and type reproduction numbers.
In addition to the approximation that cuts off the sum in Eq.~(\ref{mainresult}), Eq.~(\ref{r0b}), which is an approximation when the degree correlation itself is small, was derived.
However, the effects of loops, which cannot be ignored in networks with high-clustering coefficients, was not considered in this study.
In addition, it is necessary to introduce a correction for the difference between the SIR and SIS models, considering the infection process \cite{morita2022b}.
Finally, the proposed eigenvalue decomposition method is expected to be extensively applied to the dynamics of networks other than epidemic models.

% If you have acknowledgments, this puts in the proper section head.
\begin{acknowledgments}
This work was supported by JSPS KAKENHI (grant No. 21K03387).
Part of this study was conducted at the 
Joint Usage/Research Center on Tropical Disease, Institute of Tropical Medicine, Nagasaki University (2023-Seeds-01).
\end{acknowledgments}
\bibliography{ref2020.bib}

\end{document}